\documentclass[aps,groupedaddress,twocolumn,letterpaper]{revtex4}
\pdfoutput=1
\usepackage{graphics}
\usepackage{epsfig}
\usepackage{longtable}

\usepackage{palatino}

\usepackage{color}

\usepackage{amssymb}
\usepackage{amsmath}
\usepackage{graphicx}
\usepackage{float}

\def \be{\begin{equation}}
\def \ee{\end{equation}}
\def \bew{\begin{widetext}\begin{equation}}
\def \eew{\end{equation}\end{widetext}}
\def \bmlett{\begin{mathletters}}
\def \emlett{\end{mathletters}}

\def \r{{\bf r}}

\def\be{\begin{equation}}
\def\ee{\end{equation}}

\def\w01{\omega_{01}}
\def\r0{R_0}

\def\omegam{\omega_{\rm M}}

\begin{document}

\title{Optomechanics}

\author{Florian Marquardt}
\affiliation{Arnold Sommerfeld Center for Theoretical Physics, Center for NanoScience, and Department of Physics, Ludwig-Maximilians-Universit\"at
M\"unchen\\ Theresienstr. 37, D-80333 M\"unchen, Germany}

\author{S.M. Girvin}
\affiliation{Department of Physics, Yale University\\
PO Box 208120, New Haven, CT 06520-8120}

\date{\today}
\maketitle

The concept that electromagnetic radiation can exert forces on material objects was predicted by Maxwell, and the radiation pressure of light was first observed experimentally more than a century ago \cite{Lebedev,NicholsandHull-Science}.  The force $F$ exerted by a beam of power $P$ retro-reflecting from a mirror is $F=2P/c$.  Because the speed of light is so large, this force is typically extremely feeble but does manifest itself in special circumstances (e.g. in the tails of comets and during star formation).  Beginning in the 1970's it was appreciated that one could trap and manipulate small particles and even individual atoms with optical forces \cite{1980_Ashkin_ReviewRadiationPressure,Ashkin-Chu1986}. 

Recently there has been a great surge of interest in the application of radiation forces to manipulate the center of mass motion of mechanical oscillators covering a huge range of scales from macroscopic mirrors in the LIGO project \cite{2004_Corbitt_ReviewLIGO,2006_12_NergisMavalvala_LIGO} to nano- or micromechanical cantilevers \cite{2004_12_HoehbergerKarrai_CoolingMicroleverNature,2006_05_AspelmeyerZeilinger_SelfCoolingMirror,2006_07_Arcizet_CoolingMirror,2006_11_Bouwmeester_FeedbackCooling, 2007_Favero_OpticalCoolingMicromirror, 2008_07_Lehnert_MicrowaveNanomechanics}, vibrating microtoroids  \cite{2005_06_Vahala_SelfOscillationsCavity,2006_11_Kippenberg_RadPressureCooling} and membranes \cite{2007_07_Harris_MembraneInTheMiddle}.  Positive damping permits cooling of the motion, negative damping permits parametric amplification of small forces \cite{2004_KarraiConstanze_IEEE,2005_06_Vahala_SelfOscillationsCavity,2007_11_LudwigNeuenhahn_SelfInducedOscillations}. Cooling a mechanical system to its quantum ground state is a key goal of the new field of optomechanics.

Radiation pressure also appears in the form of unavoidable random back action forces accompanying optical measurements of position as the precision of those measurements approaches the limits set by quantum mechanics \cite{1992_BraginskyKhalili_QuantumMeasurement,2008_10_ClerkEtAl_QuantumNoiseReview}.   The randomness is due to the photon shot noise, whose observation is a second key goal of the field.

In pioneering work, Braginsky and collaborators \cite{1967_BraginskyManukin_PonderomotiveEffectsEMRadiation,1970_Braginsky_OpticalCoolingExperiment} first detected mechanical damping due to radiation in the decay of an excited oscillator.  Very recently both measurement and mechanical damping of (the much smaller) random thermal Brownian motion (i.e.~cooling of the center of mass motion) was achieved by several groups using different techniques (see also \cite{2008_08_KippenbergVahala_ScienceReview} for a brief review). These include the intrinsic optomechanical cooling (to be described below) by photothermal forces  \cite{2004_12_HoehbergerKarrai_CoolingMicroleverNature} or radiation pressure  \cite{2006_05_AspelmeyerZeilinger_SelfCoolingMirror,2006_07_Arcizet_CoolingMirror,2006_11_Kippenberg_RadPressureCooling,2006_12_NergisMavalvala_LIGO,2007_07_Harris_MembraneInTheMiddle} and active feedback cooling \cite{1999_10_Cohadon_CoolingMirrorFeedback,2006_11_Bouwmeester_FeedbackCooling} based on position measurements. 

\begin{figure}
\includegraphics[width=1.0\columnwidth]{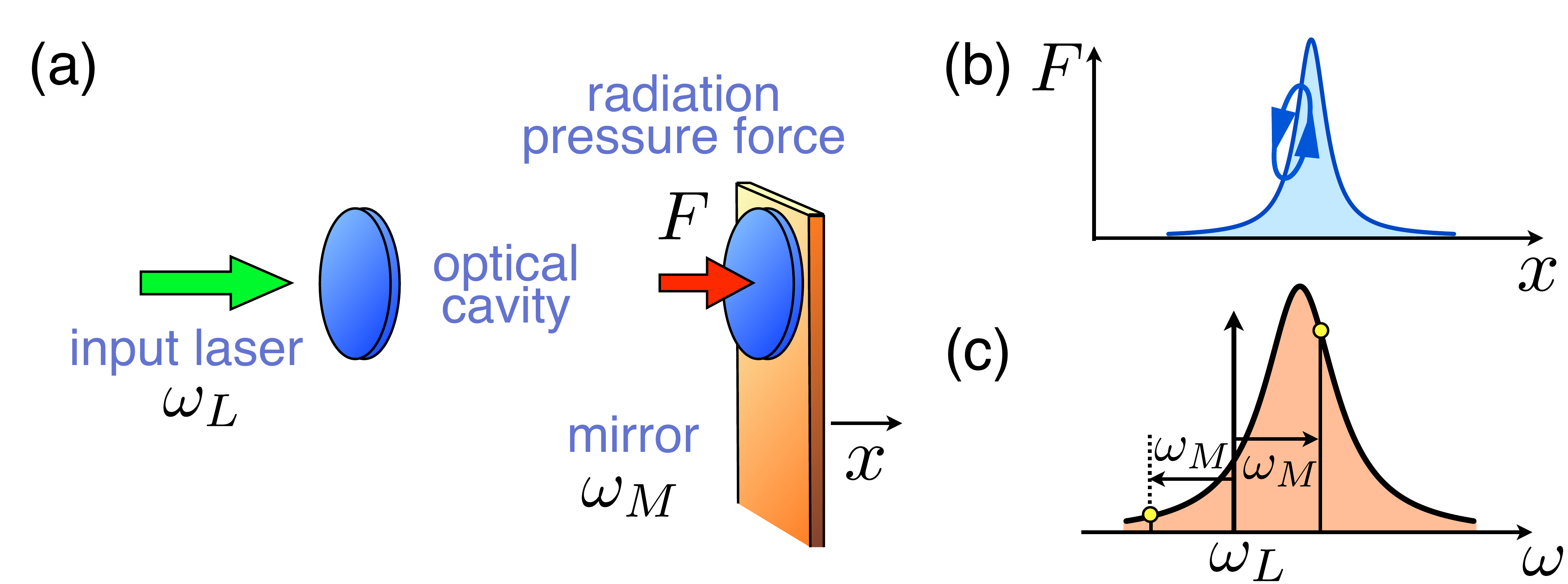}
\caption{(a) Schematic optomechanical setup. (b) Radiation pressure force vs. position. (c) In the quantum picture of cooling, Raman-scattered laser photons see a density of states that is changed by the presence of the cavity.}
\label{FigSetup}
\end{figure}

\section*{Retarded radiation forces}

The typical experimental setup in optomechanics consists of an optical cavity where one of the end-mirrors can move (Fig.~\ref{FigSetup}). For example, experimentalists have attached micromirrors to AFM cantilevers or nanobeams \cite{2006_05_AspelmeyerZeilinger_SelfCoolingMirror,2006_07_Arcizet_CoolingMirror,2006_11_Bouwmeester_FeedbackCooling}. When the cavity is illuminated by a laser, the circulating light gives rise to a radiation pressure force that deflects the mirror. Any displacement of the mirror, in turn, will change the cavity's length and thereby alter the circulating intensity, since the optical cavity mode frequency shifts with respect to the fixed laser frequency. It is this coupled dynamics that produces a wealth of interesting effects in such systems. The role of the cavity is twofold: It resonantly enhances the circulating intensity, and it makes the intensity depend very sensitively on the position. Although the setup described here may seem rather special at first sight, it is in fact just one incarnation of a very generic nonlinear nonequilibrium situation: On the most general level, we are dealing with a resonance (the optical cavity mode) that is driven (by a laser), and whose resonance frequency is pulled by the displacement of some mechanical degree of freedom (the movable mirror). Having the resonance frequency depend in this manner on the position immediately implies that there will be a mechanical force. Given this general description, it is no wonder the same physics has by now been realized in a diverse variety of physical systems, including superconducting microwave circuits \cite{2008_07_Lehnert_MicrowaveNanomechanics} and ultracold atoms \cite{2007_08_Murch_AtomsCavityHeating,2008_10_Esslinger_ColdAtomsOptomechanics}. However, in the following we will employ the terms appropriate for the simple optical setup, keeping in mind that the concepts can readily be translated to other situations.

Intrinsically, the movable mirror is a harmonic oscillator. However, as the radiation force depends on the mirror's position, it modifies the mechanical properties of the mirror. The force gradient will change the mirror's spring constant, an effect known as "optical spring", which has been used to increase the frequency of a mirror by a factor of more than twenty, essentially trapping it using light \cite{2006_12_NergisMavalvala_LIGO}. The potential in which the mirror moves can be changed drastically by the radiation forces, eventually giving rise to  multiple stable positions if the circulating intensity is large enough \cite{1983_10_DorselWalther_BistabilityMirror}.

There is yet another crucial feature about the radiation forces: they respond with a time-lag. In the setup discussed here, this is due to the finite ring-down time of the cavity, i.e. the time needed for photons to leak out (proportional to the cavity's finesse). The radiation force as a function of mirror position is a simple Lorentzian (Fabry-Perot resonance). Let us imagine that the mirror is placed on the slope of the resonance (see Fig.~\ref{FigSetup}).  As the mirror oscillates, e.g.  due to thermal fluctuations or because of driving, it moves back and forth along the slope. On approaching the resonance, the force will be smaller than expected, due to the time-lag, and it  remains larger when the mirror retracts. Overall, the radiation force extracts work from the mirror: $\oint F dx<0$. This amounts to an extra damping, which will cool down the mirror by reducing thermal fluctuations. As discussed  below, positioning the mirror on the opposite side of the resonance leads to a negative effective damping constant. These effects are sometimes labeled "dynamical back-action", since they involve the light field acting back on the mechanical motion after having been perturbed by the mirror. Alternative optomechanical cooling schemes include Doppler-cooling in Bragg mirrors \cite{2007_06_Karrai_DopplerCooling} and "active feedback cooling" \cite{1999_10_Cohadon_CoolingMirrorFeedback,2006_11_Bouwmeester_FeedbackCooling,2007_GenesAspelmeyer_CoolingPRA}. 

The optomechanical damping rate $\Gamma_{\rm opt}$ scales linearly with laser intensity and depends sensitively on the position of the mirror. In the naive classical picture described here, it reduces the effective temperature according to $T_{\rm eff}=T \Gamma / (\Gamma + \Gamma_{\rm opt})$, where $T$ is the bulk equilibrium temperature and $\Gamma$ the intrinsic mechanical damping rate. Note that we are talking about the effective temperature of a single mechanical mode of the structure that carries the mirror: Optomechanical cooling will not reduce the bulk temperature of the setup. This, however, is fully sufficient, if in the end the experiment is only sensitive to this particular degree of freedom. An analogous situation arises in molecule interferometers, where the center-of-mass motion may be quantum, even though the internal motion of atoms in the molecule remains hot.

\section*{Quantum picture of cooling: Towards the ground state}

The classical time delay description given above shows how the viscous damping force is produced.  As a transition to the full quantum picture, it is convenient to switch from the time domain to the frequency domain.  Periodic motion of the mechanical system at frequency $\omegam$ leads to amplitude and phase modulation of the optical amplitude inside the cavity.  This modulation leads to sidebands displaced from the optical carrier frequency by $\pm \omegam$.  This is precisely analogous to Raman scattering from a solid whose index of refraction is periodically modulated in time (and space) by sound waves.  Hence the lower and upper sidebands are referred to as Stokes and anti-Stokes respectively.  If both phase and amplitude modulation are present, they interfere causing one sideband to be stronger than the other.  This can be achieved by detuning the optical carrier frequency from the cavity resonance.

Quantum mechanically, the lower sideband comes from a process in which a carrier photon loses energy $\hbar\omegam$ by creating a phonon inside the mechanical oscillator.  Correspondingly the anti-Stokes upper sideband comes from a process that removes energy $\hbar\omegam$ from the mechanical oscillator.  This is the process needed for cooling. Because the sideband photons differ in energy by $2\hbar\omegam$, a difference in intensity of the two sidebands implies a net energy transfer by the optical field from or to the mechanical system.  The required asymmetry is achieved by putting the optical carrier frequency below the nominal cavity Fabry-Perot resonance.  As shown in Fig.~\ref{FigSetup}, this puts the anti-Stokes line closer to the cavity resonance and the Stokes line further away.  This yields an asymmetry in the density of states seen by the Stokes and anti-Stokes photons and hence an asymmetry in the rate of their production, as can be analyzed nicely in the "quantum noise" approach \cite{2007_01_Marquardt_CantileverCooling,2008_10_ClerkEtAl_QuantumNoiseReview}.

Although this scheme produce cooling, we cannot approach the quantum ground state unless the Stokes intensity is close to zero.  This is reasonable since the Stokes process excites the mechanical system to higher energy levels.  As shown in Fig.~(\ref{FigSetup}) the huge Stokes/anti-Stokes asymmetry can be achieved only in the good cavity limit where the cavity resonance linewidth is smaller than the sideband spacing $2\omegam$. Another condition is that the optical intensity be high enough that the resulting optical damping almost instantly removes any thermal phonons which enter the mechanical oscillator from the surroundings.  Then, the full quantum expression for the minimum achievable mean phonon number of the oscillator is \cite{2007_01_Marquardt_CantileverCooling,2007_02_WilsonRae_Cooling}

\be
{\bar n}_{\rm min} = \left( {\kappa \over 4 \omega_M } \right)^2, \label{coolinglimit}
\ee
where $\kappa$ is the optical ring-down rate of the cavity. While not technically easy, one can in principle detect the approach to the mechanical ground state by the disappearance of the anti-Stokes sideband. Mechanical and optical resonances hybridize \cite{2007_01_Marquardt_CantileverCooling,2008_03_MarquardtClerkGirvin_ReviewOptomechanics} in the strong-coupling regime when $\Gamma_{\rm opt}$ exceeds the cavity decay rate $\kappa$.

\begin{figure}
\includegraphics[width=1.0\columnwidth]{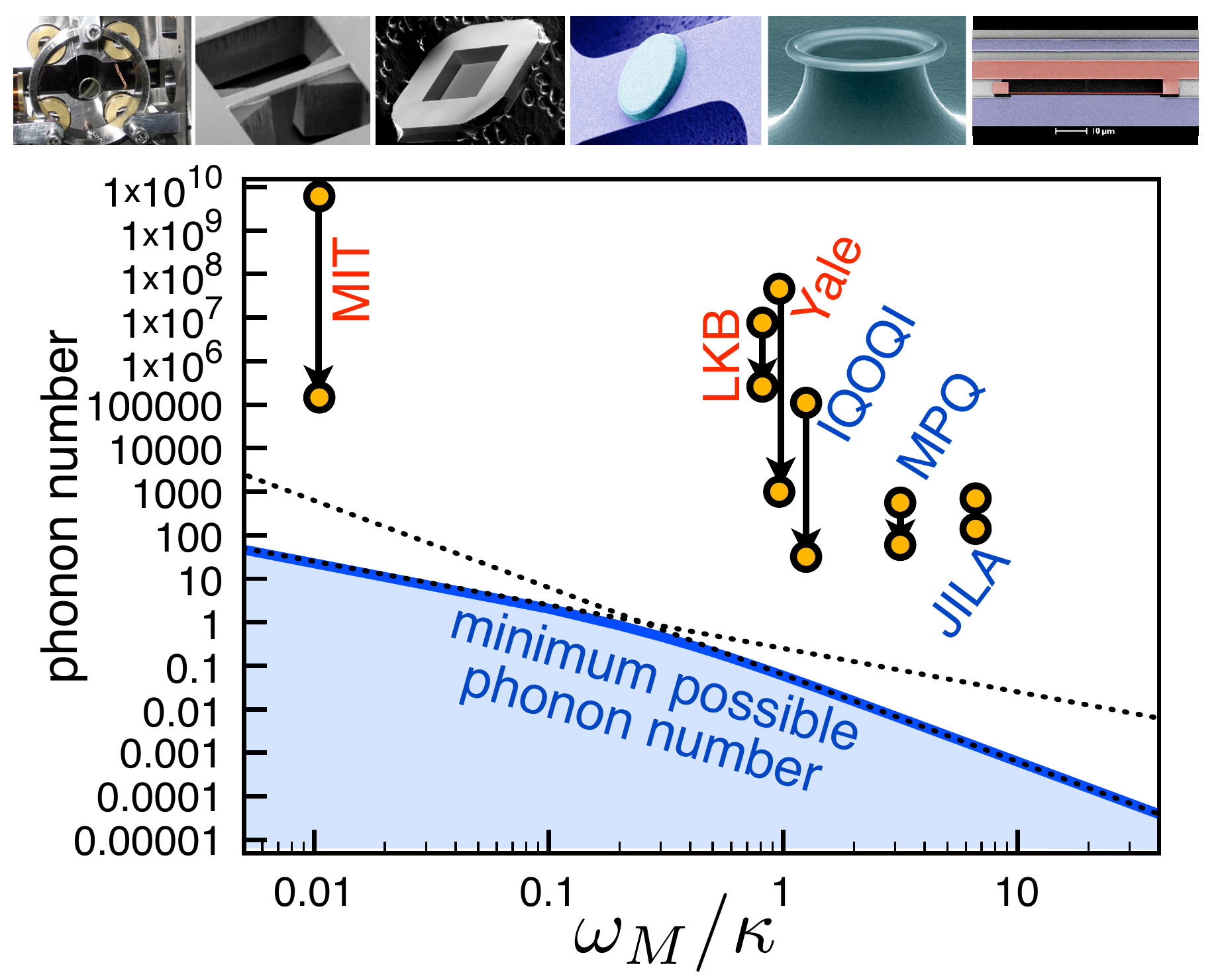}
\caption{Examples of recent progress in optomechanical cooling. The initial and final phonon numbers are plotted vs. mechanical frequency divided by the optical linewidth. The quantum limit for optomechanical cooling is indicated as a blue curve \cite{2007_01_Marquardt_CantileverCooling,2007_02_WilsonRae_Cooling}. $\omegam/\kappa\ll 1$ is the 'bad cavity' limit, and $\omegam/\kappa\gg1$ is the 'good cavity' limit, for which Eq.~\ref{coolinglimit} holds and ground-state cooling is possible. Red labels indicate cooling from room temperature, blue labels refer to cryogenic setups. Initial phonon numbers vary even for the same temperature due to different frequencies. Data (and setup pictures, left to right) from experiments at MIT \cite{2006_12_NergisMavalvala_LIGO}, Laboratoire Kastler Brossel (LKB)  \cite{2006_07_Arcizet_CoolingMirror}, Yale \cite{2007_07_Harris_MembraneInTheMiddle}, Vienna (IQOQI) \cite{2009_01_Aspelmeyer30phonons}, MPQ Munich \cite{2009_01_KippenbergFewPhonons}, and JILA at Boulder \cite{2008_11_LehnertCoolingPRL}.}
\label{OverviewCooling}
\end{figure}

At present, experiments have not yet reached the ground state, though phonon numbers as low as $30$ have been obtained very recently using optomechanical cooling \cite{2009_01_Aspelmeyer30phonons,2009_01_KippenbergFewPhonons}. Current challenges include starting from a low bulk temperature (requiring cryogenic operation), making sure to have a large mechanical quality factor (which limits the achievable cooling ratio), and fighting spurious heating from light absorption. Figure \ref{OverviewCooling} illustrates the current status for intrinsic cooling (without feedback).

\section*{Displacement readout} 

Detecting the mirror's motion is in principle straightforward, since the optical
phase shift is directly proportional to the mirror's
displacement $x$. Typically, the Lorentzian frequency spectrum of the mirror's position fluctuations is obtained in this way. The peak width
yields the total damping rate, including the effective optomechanical damping.
 The area under the spectrum reveals the variance of $x$,
which is a measure of the effective temperature, according to the classical
equipartition theorem. 

It is well known that quantum mechanics puts a fundamental constraint on the
sensitivity of any such "weak" displacement measurement\cite{1992_BraginskyKhalili_QuantumMeasurement,2008_10_ClerkEtAl_QuantumNoiseReview}. Indeed, being able to follow
the motion over time with arbitrary precision would reveal the mirror's trajectory, which
is forbidden by Heisenberg's uncertainty relation. The photon shot noise limits the precision for estimating the phase shift. In principle, this can be overcome by increasing the light intensity. However, then another effect kicks in: The shot noise of photons being reflected from the mirror imprints an unavoidable "jitter", masking the mirror's "intrinsic" motion. This effect is called measurement back-action. The standard quantum limit is reached when both effects are equally strong. It corresponds to resolving the mirror's position to within its ground state uncertainty, after averaging the signal over a damping time. The quantum limit has been approached up to a factor of five recently\cite{2009_01_KippenbergFewPhonons}, with an imprecision of $10^{-18} m/\sqrt{\rm Hz}$. Detecting the measurement back-action effects is still an outstanding challenge (but see \cite{2008_09_LKB_BeamCorrelations}). Back-action free measurements of quadratures of the mechanical motion\cite{2008_02_Clerk_Squeezing} are another option.

However, in order to see genuine "quantum jumps", it is necessary to carry out 
a quantum non-demolition measurement with respect to an observable that, unlike position,
is conserved by the Hamiltonian. The most important example in this
context would be the phonon number. Recently, a modified optomechanical
setup was introduced \cite{2007_07_Harris_MembraneInTheMiddle,2008_05_Jayich_MIMlong}, with a movable membrane in-between two fixed end-mirrors.
In such a situation, the optical frequency shift can be made to depend
quadratically on the displacement. This would enable phonon number (Fock state) detection,
once the parameters are optimized further and the system can be cooled into the
quantum regime.

\section*{Nonlinear dynamics, instability, amplification}

\begin{figure}
\includegraphics[width=1.0\columnwidth]{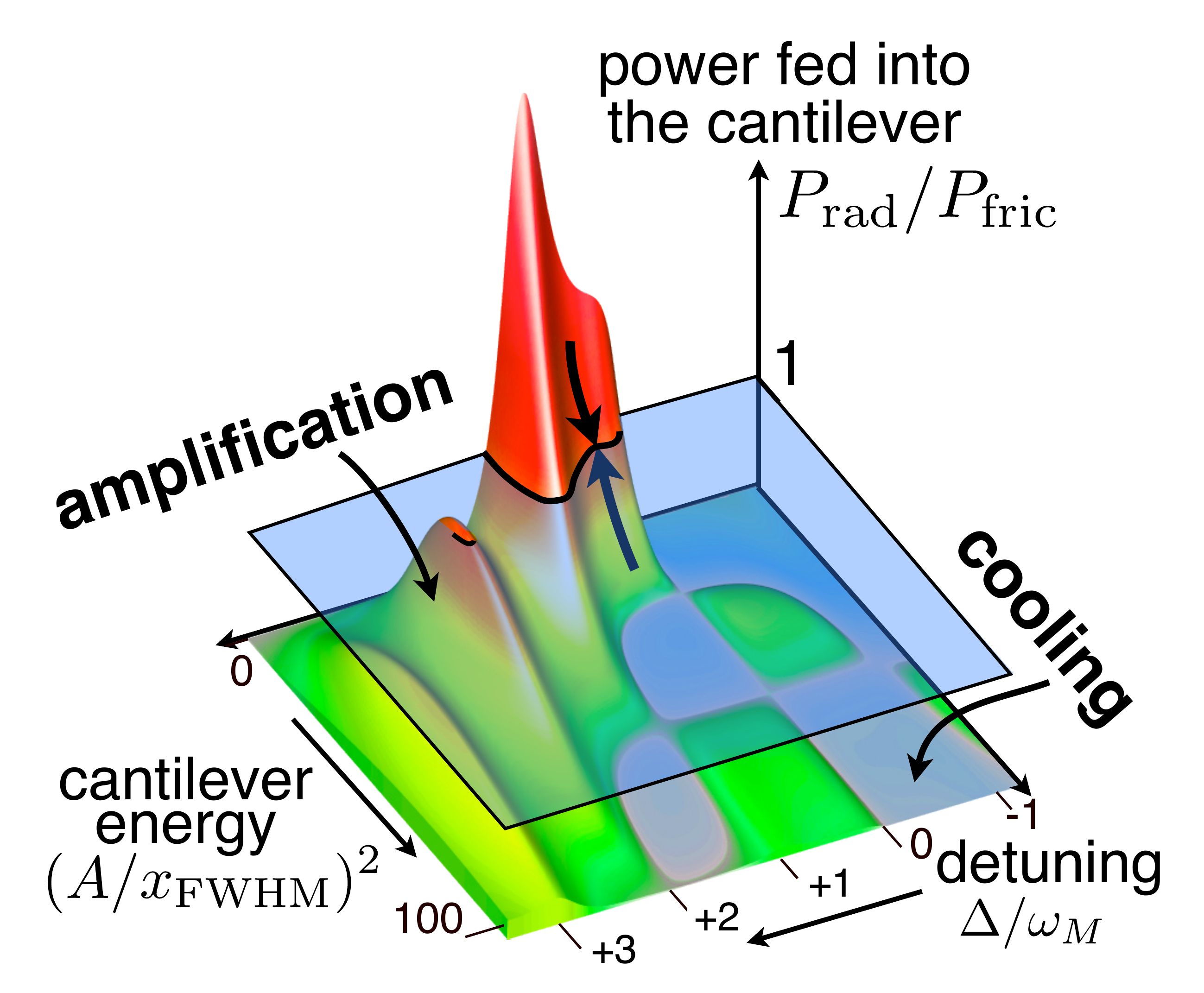}
\caption{Attractor diagram for nonlinear optomechanical motion. Stable self-sustained oscillations occur when the average power $P_{\rm rad}$ fed into the mechanical motion by the radiation pressure equals the power $P_{\rm fric}$ dissipated through friction. Their ratio depends on the amplitude $A$ of mirror motion (plotted in terms of the width of the optical resonance, $x_{\rm FWHM}$), and on the detuning between laser and optical resonance, as well as the input power (fixed in this plot).  }
\label{AttractorFigure}
\end{figure}

Beyond the linearized dynamics discussed up to now, 
such systems can display much richer, nonlinear effects as well. 
Recall that on the decreasing slope of the force vs. position curve, work is performed on the mirror, effectively reducing the overall damping as $\Gamma_{\rm opt}$  now becomes negative. Once the laser intensity is strong enough to make the total "damping rate" negative, any tiny amplitude oscillation will grow exponentially \cite{1987_10_AguirregabiriaBel_DelayInducedInstabilityFP,2001_07_Braginsky_ParametricInstabilityFPCavity,2005_02_MarquardtHarrisGirvin_Cavity}. This growth finally saturates due to nonlinear effects, and the mirror settles into periodic, self-sustained oscillations, as observed in experiments \cite{2004_KarraiConstanze_IEEE,2005_06_Vahala_SelfOscillationsCavity,2007_11_LudwigNeuenhahn_SelfInducedOscillations}. Their amplitude is determined by the laser intensity, the detuning, and the strength of the intrinsic mechanical friction, as well as other parameters.

Note that the parametric instability we have just described is conceptually identical to what happens in a laser above the lasing threshold. Here, the mechanical vibration plays the role of the laser's light mode, and the pump is provided by the radiation that drives the cavity. 

To obtain the attractors for the motion, one may pose a simple question: How does the work per cycle performed by the radiation field depend on the mirror's oscillation amplitude? The power fed into the system has to match the power dissipated by friction. When one draws a map of the possible amplitudes of oscillation that are consistent with this condition, an intricate structure emerges \cite{2005_02_MarquardtHarrisGirvin_Cavity,2008_03_Luwig_OptoInstabilityQuantum} (see Fig.~\ref{AttractorFigure}). In particular, at fixed parameters a large number of possible amplitudes may exist simultaneously. This multistability has begun to be explored in experiments \cite{2007_11_LudwigNeuenhahn_SelfInducedOscillations}, and it might even be useful for sensitive measurements \cite{2005_02_MarquardtHarrisGirvin_Cavity}. 

At even higher optical drive powers, the mirror may enter a state of chaotic motion \cite{2007_Carmon_ChaosPRL}, which still remains mostly unexplored. In addition, one may ask about possible quantum effects in the nonlinear dynamics \cite{2008_03_Luwig_OptoInstabilityQuantum}. 

\section*{Nonclassical states, squeezing, entanglement}

The question arises how to use the optomechanical interaction to produce genuinely nonclassical states of the light field and/or the mechanical motion. We list some ideas in the following that may be implemented in the future. 

As we have seen, the cavity length changes in response to the circulating intensity. In this regard, the setup is equivalent to a nonlinear optical medium, with an intensity-dependent index of refraction. Such a Kerr medium may be used to produce squeezing in the light field, e.g. by suppressing the intensity fluctuations (amplitude squeezing), and this can be translated directly to optomechanics \cite{1994_02_Fabre_SqueezingInCavityWithMovableMirror}. With regard to the mirror, squeezed states might be produced by varying the optical spring constant in time. As indicated above, mechanical Fock states could be produced via measurements.  

Entanglement between the light field and the mirror can be generated easily, in principle. Suppose for a moment that the cavity is closed and the field is in a superposition of different photon numbers, e.g. in a coherent state. Each of these Fock states of the radiation field will exert a different radiation pressure force, thereby displacing the mirror by a different amount. This creates an entangled state, which may be called a "Schr\"odinger cat", as the mirror involves many billions of atoms (see e.g. \cite{1997_04_Mancini_MirrorCat,1999_05_Bose_Cat,2003_09_Marshall_QSuperposMirror, 2007_Aspelmeyer_EntanglementLightMirror} and others). Remarkably, after a full period of the mirror oscillation, the entanglement would be undone, like in a quantum eraser experiment. 
It has been suggested that producing entanglement in this way and checking for its decay over time could eventually be a means to test for potentially unknown sources of decoherence, probably even including hypothetical gravitationally induced collapse of the wave function of the massive mirror \cite{2003_09_Marshall_QSuperposMirror}. 

When several movable mirrors or membranes are included, the radiation field can be exploited as a medium which couples these mechanical elements to each other \cite{2007_06_VitaliManciniTombesi_EntanglementMirrors,2008_11_HartmannPlenio_EntanglementMembranes,2008_Meystre_MultiMembraneEntanglement}, leading to entanglement if thermal fluctuations are sufficiently suppressed. Experimental proof of entanglement then requires correlation measurements via optical probe beams.

\section*{Overview of experimental setups}

Among the setups that have been realized during the past five years, most
involve cantilevers \cite{2004_12_HoehbergerKarrai_CoolingMicroleverNature,2006_11_Bouwmeester_FeedbackCooling} or nanobeams \cite{2006_05_AspelmeyerZeilinger_SelfCoolingMirror,2006_07_Arcizet_CoolingMirror} as mechanical elements. Masses typically range from $10^{-15} {\rm kg}$ to $10^{-10} {\rm kg}$ (and even $1 {\rm g}$ \cite{2006_12_NergisMavalvala_LIGO}), while frequencies are often in the MHz regime ($\omegam/2\pi=1 {\rm kHz}$ to $100 {\rm MHz}$). Light is typically reflected from Bragg mirrors made from multi-layered dielectric materials. A rather different approach is based on microtoroid optical cavities made from silica on a chip \cite{2005_06_Vahala_SelfOscillationsCavity,2006_11_Kippenberg_RadPressureCooling,2007_03_Carmon_ToroidModesPRL}. The light circulating inside an optical whispering gallery mode inside the toroid exerts a radiation pressure that couples to a mechanical breathing mode.

The biggest challenge in all of these devices is to obtain both a high optical finesse (currently in the range from $10^3$ to $10^5$), and a high mechanical quality factor ($10^3$ to $10^5$ for beams and cantilevers).
As explained above, an alternative approach \cite{2007_07_Harris_MembraneInTheMiddle,2008_05_Jayich_MIMlong} involving a $50$ nanometer thin membrane inside a fixed optical cavity can circumvent this problem to some degree, and has reached a finesse of $10^4$ and a mechanical quality factor of $10^6$. 

Optomechanical ideas have recently been realized in a number of other systems as well. For example, it is possible to replace the optical cavity by driven radio-frequency \cite{2007_Wineland_RFcircuitCooling} or microwave \cite{2008_07_Lehnert_MicrowaveNanomechanics,2008_11_LehnertCoolingPRL} circuits, whose resonance frequency depends on the motion of a capacitively coupled nanobeam. The setup involving superconducting microwave resonators is especially promising as it can be coupled to Josephson junctions, qubits and amplifiers on the same chip. Incidentally, the essence of optomechanical cooling has also been demonstrated using a current-driven superconducting single electron transistor in place of the optical cavity \cite{2006_08_Schwab_CPB_Molasses}.

Another recent development exploits the radiation forces that occur between two glass fibres or between a fibre and a dielectric substrate, where the coupling is through the evanescent light field \cite{2007_07_Painter_MicroDiskWaveGuide_Optomechanics,2008_11_HongTang_PhotonicCircuitMechanics}. These devices operate on the nanoscale, and they can generate large forces without the need for a high finesse cavity. One may thus envisage integrating mechanical devices with photonic crystals, fibres and other optical elements on a chip, serving as the basis for optically controlled mechanical information processing and sensing.

For a long time, radiation forces had already been used to cool, trap and manipulate atoms, before being applied to mechanical structures. It is therefore amusing to note that the concepts of optomechanics are being transferred back to the domain of cold atoms. Several experiments \cite{2007_08_Murch_AtomsCavityHeating,2008_10_Esslinger_ColdAtomsOptomechanics} have now demonstrated how the mechanical motion of clouds of ultracold atoms inside an optical cavity can couple to the light field and display the effects we have been discussing. Given the small mass of the atom cloud, the mechanical effects of a single photon can be significant. This allows to study optomechanics in a new domain. One might also entangle an atomic ensemble and a nanomechanical system (e.g. \cite{2009_01_Hammerer_NanomechanicsAtomEnsembles_PRL}). 

\section*{Outlook, new directions, and challenges}

In the short term, experiments are racing towards the ground state of mechanical motion, to enable manipulation in the quantum regime. Achieving this goal would open the door towards possible applications, for example in the area of quantum information processing. It would also permit us to answer fundamental questions, such as whether we understand decoherence processes in massive objects. Sensitive measurements (of displacement, mass, etc.) are another area where optomechanical systems will find applications, and while they do not urgently require going into the quantum regime, they could benefit from the improved sensitivity. 

In the longer term, optomechanics may also be viewed as a light-mechanics interface to realize hybrid structures for (classical or quantum) information processing, switching or storage, in integrated photonic circuits on a semiconductor chip. 

We acknowledge support by (S.M.G.) NSF grants DMR-0653377 and DMR-0603369, as well as (F.M.) the Emmy-Noether program, NIM, and SFB 631.

\bibliography{optomechreferencesNew}

\end{document}